\documentclass{epl}

\title{Gaussian density fluctuations, mode coupling theory, and all that}
\author{Grzegorz Szamel}
\institute{
Department of Chemistry, Colorado State University,
Ft. Collins, CO 80523, USA }
 
\begin{document}
 
\maketitle
 
\begin{abstract}
We consider a toy model for glassy dynamics of colloidal suspensions:
a single Brownian particle diffusing among immobile obstacles.
If Gaussian factorization of \emph{static} density 
fluctuations is assumed,
this model can be solved without factorization approximation for 
any \emph{dynamic} correlation function. The solution differs from that
obtained from the ideal mode coupling theory (MCT). 
The latter is equivalent to including only some, positive
definite terms in an expression for the memory function. 
An approximate re-summation of the complete expression
suggests that, under the assumption of Gaussian factorization of 
static fluctuations, 
mobile particle's motion is always diffusive. 
In contrast, MCT predicts that the mobile particle becomes 
localized at a high enough obstacle density. We discuss
the implications of these results for models for glassy dynamics. 
\end{abstract}

\section{Introduction}
During the last decade considerable effort has been devoted to
simulational and experimental verification of the mode coupling theory (MCT)
of glassy dynamics and the glass transition \cite{KobLH,Goetze,SlowRel}. 
The consensus that emerged 
from this work is that MCT describes in a satisfactory way
``weakly'' supercooled liquids (\textit{i.e.} it describes 
the first few decades of slowing down on approaching the
glass transition). In particular, MCT has been quite successful
when applied to concentrated colloidal suspensions \cite{review},
the colloidal glass \cite{colglass}, 
and gelation \cite{colgel} transitions.

Notably, less effort has been devoted to the foundations 
of the mode coupling theory (see, however, Refs. \cite{Oppenheim,Dawson,HCA}).
This is somewhat surprising in view of MCT's 
several well-known problems. The most important, fundamental problem
is the uncontrolled nature of the basic MCT approximation: factorization
of a complicated \emph{time-dependent} pair-density (\textit{i.e.}
four-particle) correlation function.

Recently, we proposed an extension of MCT for dynamics of colloidal
suspensions and the colloidal glass transition \cite{ownPRL}. 
Our theory includes, in an approximate way, time-dependent pair-density
fluctuations. It relies upon a factorization approximation that is 
similar to that used in MCT, but is applied at a level of a memory 
function for the time-dependent pair-density correlation function. 
The theory predicts an ergodicity breaking transition similar
to that of MCT, but at a higher density. Thus it partially solves another
well-known MCT problem: overestimation of so-called dynamic feedback 
effect and the resulting underestimation of the colloidal glass transition
density.

Here, for a simpler, toy model, we go further: we completely 
avoid using factorization approximation for any \emph{dynamic} correlation
function. We only assume Gaussian factorization of \emph{static}
correlations \cite{Gausfac}. 
It should be noted that a frequently used approach
to glassy dynamics is to start from a set of fluctuating
hydrodynamics equations which are supplemented by a 
quadratic free energy
implying Gaussian static density fluctuations \cite{DR,Kawasaki}. 
We argue that the 
analysis presented here has implications for such models.

Since the approach is technically quite involved, we  
state the main results immediately: we derive an essentially exact 
expression for the time-integrated memory function for a single 
Brownian particle moving among immobile obstacles. We compare 
this expression with one derived from MCT and show that the latter 
includes a subset of the former's terms: only 
explicitly positive terms from the exact series (\textit{i.e.} the terms 
that always increase the effective friction felt
by the mobile particle) are included within MCT. 
This is the origin of MCT's overestimation of 
the dynamic feedback effect. 
An approximate re-summation of the exact series suggests that,
under the assumption of Gaussian static fluctuations,  
the mobile particle's motion is always diffusive. 
In contrast, MCT predicts that the mobile particle becomes 
localized at high enough obstacle density. 

This result has important consequences for models used to 
study glassy dynamics. We show here that, 
if static correlations are Gaussian, a single mobile particle is never
localized by immobile obstacles. This suggests that  
a similar \emph{fully mobile}
system (\textit{i.e.} Gaussian static 
correlations and all particles diffusing) cannot
undergo an ergodicity breaking transition. In other words, the ergodicity
breaking transition predicted for such a system by a mode coupling theory
is, most probably, an artifact of the factorization approximation. 

Note that this does \emph{not} mean that MCT is qualitatively wrong
for a system with complicated many-particle static correlations (like,
\textit{e.g.}, the hard sphere system). It can be argued that terms
that cut-off MCT's localization transition 
(\textit{i.e.} terms that are neglected in MCT)
are canceled by other terms that originate from non-Gaussian
static correlations. Indeed, empirical success of MCT for colloidal
systems suggests that this might be the case. 
It is at present unclear how to describe this 
remarkable cancellation.

\section{Toy model}
We consider one spherical Brownian particle diffusing between $N-1$ immobile,
spherically symmetric obstacles. The particle interacts with the obstacles 
via a potential $V(r)$. The obstacles are mechanically identical to 
the mobile particle. We assume that the initial joint probability
distribution for the mobile particle and the obstacles is given
by the equilibrium canonical distribution at temperature $T=(k_B\beta)^{-1}$.
The time evolution of the system is described by a  
generalized Smoluchowski equation:
\begin{eqnarray}\label{NSm}\nonumber
\frac{\partial}{\partial t} P_N(\mathbf{r}_1 | \mathbf{r}_2, ...,
\mathbf{r}_N; t) &=& D_0 \frac{\partial}{\partial\mathbf{r}_1} \cdot
\left(\frac{\partial}{\partial\mathbf{r}_1} - \beta
\mathbf{F}_{1}\right) P_N(\mathbf{r}_1 | \mathbf{r}_2, ...,
\mathbf{r}_N; t) \\ &=& \Omega P_N(\mathbf{r}_1 | \mathbf{r}_2, ...,
\mathbf{r}_N; t)
\end{eqnarray}
with the initial condition
\begin{equation}\label{NSm0}
P_N(\mathbf{r}_1 | \mathbf{r}_2, ...,
\mathbf{r}_N; t=0) =  P_N^{eq}(\mathbf{r}_1,\mathbf{r}_2, ...,
\mathbf{r}_N)\delta(\mathbf{r}_1-\mathbf{r}_0).
\end{equation}
Here $\mathbf{r}_1$ denotes the position of the mobile particle
and $\mathbf{r}_2, ..., \mathbf{r}_N$ denote positions of the obstacles.
Furthermore, $D_0$ is the diffusion coefficient of the mobile particle
in the absence of the obstacles, and 
$\mathbf{F}_1 = \sum_{j>1} \mathbf{F}_{1j} = -\sum_{j>1}\mathbf{\nabla}_1
V(r_{1j})$ is the force acting on it.
Finally, the second line in Eq. (\ref{NSm}) defines the $N$-particle
generalized Smoluchowski operator $\Omega$.

Our theory starts from the memory function representation of the mobile 
particle density correlation function, $F_1(k;t)$,
\begin{equation}\label{Fktdef}
F_1(k;t) = \left< n_1(\mathbf{k}) \exp(\Omega t)n_1(-\mathbf{k})
\right>.
\end{equation}
Here $n_1(\mathbf{k})$ is the Fourier transform of the mobile particle 
density,
$
n_1(\mathbf{k}) = e^{-i\mathbf{k}\cdot\mathbf{r}_1}, 
$
and $\left<\dots\right>$ denotes the average over the 
equilibrium probability distribution, $P_N^{eq}$. 
Note the equilibrium distribution
stands to the right of the quantity being averaged, and all operators
act on it as well as on everything else. 

To derive the memory function representation we start from an expression
for the Laplace transform, $LT$, of the time derivative of $F_1$:
\begin{equation}\label{Fktder}
LT(\dot{F}_1(k;t)) = zF_1(k,z) - F_1(k;t=0) = 
\left< n_1(\mathbf{k}) \Omega \frac{1}{z-\Omega} n_1(-\mathbf{k})
\right>.
\end{equation}
Using standard projection operator manipulations we rewrite
(\ref{Fktder}) in the following form:
\begin{equation}\label{Fktmr}
LT(\dot{F}_1(k;t)) = -D_0 \mathbf{k}\cdot\left(
1 - D_0^{-1} \left< \mathbf{j}_1(\mathbf{k}) 
\frac{1}{z-\hat{Q}_1\Omega\hat{Q}_1} \mathbf{j}_1 (-\mathbf{k})
\right> \right)\cdot\mathbf{k}\; F_1(k,z).
\end{equation}
Here $\mathbf{j}_1$ is a projected current density of the mobile particle,
\begin{equation}\label{j1}
\mathbf{j}(\mathbf{k}) = \hat{Q}_1 D_0 
(-i\mathbf{k} + \beta\mathbf{F}_1)e^{-i\mathbf{k}\cdot\mathbf{r}_1},
\end{equation}
$\hat{Q}_1=1-\hat{P}_1$, and $\hat{P}_1$ is a projection operator
on the mobile particle density subspace,
\begin{equation}\label{P1}
\hat{P}_1 = 
\sum_{\mathbf{q}} \cdots
n_1(-\mathbf{q})\left>\right<n_1(\mathbf{q})\cdots \equiv 
\cdots n_1(-\mathbf{q})\left>\right<n_1(\mathbf{q})\cdots .
\end{equation}
Note that in Eq. (\ref{P1}) we introduced a summation convention:
we sum over all repeated wavevectors appearing in adjacent 
\textit{ket}, $>$, and \textit{bra}, $<$.

Next we define the one-particle irreducible evolution operator,
\begin{equation}\label{O1}
\Omega_1^{irr} = \hat{Q}_1 D_0
\frac{\partial}{\partial\mathbf{r}_1} \hat{Q}_1
\cdot
\left(\frac{\partial}{\partial\mathbf{r}_1}
-\beta\mathbf{F}_1\right)\hat{Q}_1
\end{equation}
and we use the same standard projection operator manipulations to 
re-write the current-current correlation function appearing in Eq. 
(\ref{Fktmr}) in the following form:
\begin{eqnarray}\label{Fktmr2}
&& \left< \mathbf{j}_1(\mathbf{k})  
\frac{1}{z-\hat{Q}_1\Omega\hat{Q}_1}\mathbf{j}_1 (-\mathbf{k})
\right>  =
\left< \mathbf{j}_1(\mathbf{k}) 
\frac{1}{z-\Omega_1^{irr}}\mathbf{j}_1 (-\mathbf{k})
\right> \\ \nonumber &&+ 
D_0^{-1} \left< \mathbf{j}_1(\mathbf{k})
\frac{1}{z-\Omega_1^{irr}}\mathbf{j}_1 (-\mathbf{q})
\right>\cdot\left<\mathbf{j}_1(\mathbf{q})  
\frac{1}{z-\hat{Q}_1\Omega\hat{Q}_1}\mathbf{j}_1 (-\mathbf{k})
\right>
\end{eqnarray}
Combining Eqs. (\ref{Fktmr}) and (\ref{Fktmr2}) we derive the 
memory function representation for $F_1(k;z)$:
\begin{equation}\label{Fktm}
LT(\dot{F}_1(k;t)) =  -D_0
\mathbf{k}\cdot\left(\mathsf{1} + D_0^{-1} \left< \mathbf{j}_1(\mathbf{k}) 
\frac{1}{z-\Omega_1^{irr}}\mathbf{j}_1 (-\mathbf{k})
\right>\right)^{-1} \cdot\mathbf{k}\; F_1(k;z)
\end{equation}
Comparing Eq. (\ref{Fktm}) with the standard form of the memory
function representation, 
$
F_1(k;z) = 1/(z+1/(1+M_1^{irr}(k;z))),
$
we identify the irreducible memory function:
\begin{equation}\label{M1irr}
M_1^{irr}(k;z) = D_0^{-1} \hat{\mathbf{k}}\cdot\left< \mathbf{j}_1(\mathbf{k}) 
\frac{1}{z-\Omega_1^{irr}}\mathbf{j}_1 (-\mathbf{k})
\right>\cdot\hat{\mathbf{k}},
\end{equation}
where $\hat{\mathbf{k}} = \mathbf{k}/k$. 

Next we define $n_2$: the
part of the joint density of the mobile particle and the obstacles
that is orthogonal to the mobile particle's density,
\begin{equation}
n_2(\mathbf{q}_1,\mathbf{q}_2) = \hat{Q}_1 
\sum_{j> 1}e^{-i\mathbf{q}_1\cdot\mathbf{r}_1-
i\mathbf{q}_2\cdot\mathbf{r}_j}.
\end{equation}
We use the following identity that is exact for two-particle
additive interactions:
\begin{equation}\label{vertex1}
\left. \mathbf{j}_1 (-\mathbf{k}) \right> =  
\left. n_2(-\mathbf{q}_1,-\mathbf{q}_2)\right>
\left<n_2(\mathbf{q}_1,\mathbf{q}_2) n_2(-\mathbf{q}_3,-\mathbf{q}_4)
\right>^{-1}
\left<n_2(\mathbf{q}_3,\mathbf{q}_4) \mathbf{j}_1 (-\mathbf{k}) \right>,
\end{equation}
where
$\left<n_2(\mathbf{q}_1,\mathbf{q}_2) n_2(-\mathbf{q}_3,-\mathbf{q}_4)
\right>^{-1}$ denotes the kernel of the inverse integral operator.
Using Eq. (\ref{vertex1}) and an analogous identity for 
$\left< \mathbf{j}_1 (\mathbf{k}) \right.$
we can express the irreducible memory function (\ref{M1irr}) in terms of the
pair-density correlation function:
\begin{equation}\label{F2kt}
F_2(\mathbf{q}_1,\mathbf{q}_2;\mathbf{q}_3,\mathbf{q}_4;t) = 
\left<n_2(\mathbf{q}_1,\mathbf{q}_2) \exp(\Omega_1^{irr}t)
n_2(-\mathbf{q}_3,-\mathbf{q}_4)\right>.
\end{equation}
In turn, this correlation function can be subjected to the
same manipulations as the mobile particle density
correlation function $F_1(k;t)$. 
Re-tracing the steps between Eqs. (\ref{Fktdef})
and (\ref{Fktm}) we arrive at the following:
\begin{eqnarray}\label{F2ktm2}\nonumber
&& 
LT(\dot{F}_2(\mathbf{q}_1,\mathbf{q}_2;\mathbf{q}_3,\mathbf{q}_4;t)) 
= 
-D_0 \mathbf{q}_1\cdot\left(
\left<n_2(\mathbf{q}_1,\mathbf{q}_2) 
n_2(-\mathbf{q}_7,-\mathbf{q}_8)\right>^{-1} \right. \\ 
&& \left.
+ \left<n_2(\mathbf{q}_1,\mathbf{q}_2)  
n_2(-\mathbf{q}_3,-\mathbf{q}_4)\right>^{-1}
\left< \mathbf{j}_2(\mathbf{q}_3,\mathbf{q}_4) 
\frac{1}{z-\Omega_2^{irr}}\mathbf{j}_2(-\mathbf{q}_5,-\mathbf{q}_6)
\right>
\right. \\ \nonumber  && \left. \times
\left<n_2(\mathbf{q}_5,\mathbf{q}_6) 
n_2(-\mathbf{q}_7,-\mathbf{q}_8)\right>^{-1}\right)^{-1}\cdot\mathbf{q}_7 
\left< n_2(\mathbf{q}_7,\mathbf{q}_8) 
n_2(-\mathbf{q}_9,-\mathbf{q}_{10})\right>^{-1}\;
F_2(\mathbf{q}_9,\mathbf{q}_{10};\mathbf{q}_3,\mathbf{q}_4;z).
\end{eqnarray}
Here $\mathbf{j}_2$ is the two-particle projected current density,
\begin{equation}\label{j2}
\mathbf{j}_2(\mathbf{q}_1,\mathbf{q}_2) = 
\hat{Q}_2 D_0 \left(\frac{\partial}{\partial\mathbf{r}_1} + \beta\mathbf{F}_1
\right)\hat{Q}_1 \sum_{j>1} e^{-i\mathbf{q}_1\cdot\mathbf{r}_1-
i\mathbf{q}_2\cdot\mathbf{r}_2},
\end{equation}
and $\Omega_2^{irr}$ is the two-particle irreducible evolution operator,
\begin{equation}\label{O2}
\Omega_2^{irr} = \hat{Q}_2 D_0
\frac{\partial}{\partial\mathbf{r}_1} \hat{Q}_2
\cdot
\left(\frac{\partial}{\partial\mathbf{r}_1}
-\beta\mathbf{F}_1\right)\hat{Q}_2
\end{equation}
In Eqs. (\ref{j2}--\ref{O2}) 
$\hat{Q}_2=1-\hat{P}_1-\hat{P}_2$, with $\hat{P}_1$ defined by 
Eq. (\ref{P1}), and $\hat{P}_2$ being projection
operator on $n_2$:
\begin{equation}
\hat{P}_2 = \cdots
n_2(-\mathbf{q}_1,-\mathbf{q}_2)\left>
\left<n_2(\mathbf{q}_1,\mathbf{q}_2)n_2(-\mathbf{q}_3,-\mathbf{q}_4)
\right>^{-1}
\right<n_2(\mathbf{q}_3,\mathbf{q}_4)\cdots .
\end{equation}

Combining Eqs. (\ref{M1irr}), (\ref{vertex1}) and (\ref{F2ktm2}) 
we can obtain an expression for the memory function
in terms of the autocorrelation function of the 
two-particle projected current density:
\begin{eqnarray}\label{M1irr2}\nonumber
&&M_1^{irr}(k;z=0) = D_0^{-1} 
\hat{\mathbf{k}}\cdot\left< \mathbf{j}_1(\mathbf{k}) 
n_2(-\mathbf{q}_1,-\mathbf{q}_2) \right> q_1^{-2} \; 
\mathbf{q}_1\cdot 
\left<n_2(\mathbf{q}_1,\mathbf{q}_2) 
n_2(-\mathbf{q}_3,-\mathbf{q}_4)\right>^{-1} 
\\ \nonumber 
&& \times \left(
\left< n_2(\mathbf{q}_3,\mathbf{q}_4)
n_2(-\mathbf{q}_5,-\mathbf{q}_6)\right>
+ D_0^{-1} 
\left< \mathbf{j}_2(\mathbf{q}_3,\mathbf{q}_4) 
\frac{1}{z-\Omega_2^{irr}}\mathbf{j}_2(-\mathbf{q}_5,-\mathbf{q}_6)\right>
\right) 
\\ && \times 
\left<n_2(\mathbf{q}_5,\mathbf{q}_6) 
n_2(-\mathbf{q}_7,-\mathbf{q}_8)\right>^{-1} \cdot\mathbf{q}_7 
q_7^{-2} \left< n_2(\mathbf{q}_7,\mathbf{q}_8) 
\mathbf{j}_1 (-\mathbf{k})
\right>\cdot\hat{\mathbf{k}}
\end{eqnarray}
Note that to get expression (\ref{M1irr2}) we resorted to a technical
approximation that is similar to the first Enskog approximation
used to approximately invert the Boltzmann collision operator \cite{Resibois}.
This minor, technical approximation is not required in one
dimension where Eq. (\ref{M1irr2}) is exact.

It is clear this procedure can be continued \textit{ad infinitum}:
the two-particle projected current correlation
function can be expressed in terms of
the three-body density correlation function, \textit{etc.}
(note that for higher order densities, $n_m$, $m\ge 3$, 
it is advantageous to use ordered multiplets of 
wavevectors, $\mathbf{q}_2 < ... < \mathbf{q}_m$).
This is somewhat akin to the well-known
continuous fraction expansion. It is different from it in that
at each step a new function depending on a larger number of variables
is introduced. 

The resulting expressions simplify greatly if we assume Gaussian
factorization of static correlations. Note that operators $\hat{Q}_m$ 
remove contributions involving fewer than $m+1$ independent connections
between two groups of particles \cite{BC}. Thus, for example, for the
correlations of the projected $m$-particle density we get
\begin{equation}\label{nmnmG}
\left<n_m(\mathbf{q}_1, ... , \mathbf{q}_m) 
n_m(-\mathbf{k}_1, ... , -\mathbf{k}_m)\right> = 
N^{m-1}S(q_2)...S(q_m)\delta_{\mathbf{q}_1,\mathbf{k}_1} ... 
\delta_{\mathbf{q}_m,\mathbf{k}_m}.
\end{equation}
Furthermore, for the density-current correlations of the type (\ref{vertex1}) 
we get
\begin{eqnarray}\nonumber 
&& \left<n_m(\mathbf{q}_1, ... , \mathbf{q}_m) 
\mathbf{j}_{m-1}(-\mathbf{k}_1, ... , -\mathbf{k}_{m-1})\right> = 
i n N^{m-2} \left(\mathbf{q}_1-\mathbf{k}_1\right) S(q_2) ... S(q_{m})
\\ && \times \sum_{j=2}^m c(q_j) 
\delta_{\mathbf{q}_1+\mathbf{q}_j,\mathbf{k}_1}
\delta_{\mathbf{q}_2,\mathbf{k}_2}
... \delta_{\mathbf{q}_{j-1},\mathbf{k}_{j-1}}
\delta_{\mathbf{q}_{j+1},\mathbf{k}_j}...
\delta_{\mathbf{q}_m,\mathbf{k}_{m-1}},
\end{eqnarray}
where $n$ is the number density, $n=N/V$, and 
$c(q)$ is the direct correlation function, $c(q)=(S(q)-1)/(nS(q))$.

Using the above described procedure, 
under the assumption of static Gaussian fluctuations, 
we get the following expression for the
time-integrated (\textit{i.e.} $z=0$) memory function at $k=0$
\begin{eqnarray}\label{Mseries}\nonumber
&&M_1^{irr}(k=0;z=0) = \frac{n}{V}\sum_{\mathbf{q}_1} 
(\hat{\mathbf{k}}\cdot\mathbf{q}_1)^2
\frac{c(q_1)S(q_1)c(q_1)}{q_1^2}  \\ \nonumber 
&& + \frac{1}{2}\left(\frac{n}{V}\right)^2 \sum_{\mathbf{q}_1,\mathbf{q}_2}
\left(\hat{\mathbf{k}}\cdot\mathbf{q}_1\frac{1}{q_1^2}
\mathbf{q}_1\cdot\mathbf{q}_2+
\hat{\mathbf{k}}\cdot\mathbf{q}_2\frac{1}{q_2^2}
\mathbf{q}_2\cdot\mathbf{q}_1\right)^2
\frac{c(q_1)S(q_1)c(q_1)c(q_2)S(q_2)c(q_2)}{|\mathbf{q}_1+\mathbf{q}_2|^2} 
\\ \nonumber 
&& + \frac{1}{3!}\left(\frac{n}{V}\right)^3
\sum_{\mathbf{q}_1,\mathbf{q}_2,\mathbf{q}_3}
\left(\hat{\mathbf{k}}\cdot\mathbf{q}_1\frac{1}{q_1^2}
\mathbf{q}_1\cdot\mathbf{q}_2
\frac{1}{|\mathbf{q}_1+\mathbf{q}_2|^2}(\mathbf{q}_1+\mathbf{q}_2)
\cdot\mathbf{q}_3
+\textrm{perm.}\right)^2\\ &&
\;\;\;\;\;\;\;\;\;\;\;\;\;\;\;\;\;\;\;\;\;\;\;\;\;\;\;\;\;\;\;\;\;\;\;         
\times
\frac{c(q_1)S(q_1)c(q_1)c(q_2)S(q_2)c(q_2)c(q_3)S(q_3)c(q_3)}
{|\mathbf{q}_1+\mathbf{q}_2+\mathbf{q}_3|^2}  + \;\;\; ...,
\end{eqnarray}
where $\textrm{perm.}$ denotes all permutations of the wavevectors'
indices. 
Eq. (\ref{Mseries}) is the main result of this note. 
Under the assumption of static Gaussian correlations, this expression
is essentially exact. The only, technical approximation is the one
similar to the first Enskog approximation.

\section{Comparison with MCT} 
Expression (\ref{Mseries}) should be compared with that resulting
from the mode coupling theory. 
The latter approach starts from expression (\ref{M1irr}) for the 
memory function, rewrites it in terms of the pair density
correlation function using (\ref{vertex1}) and (\ref{F2kt}), and then 
resorts to a factorization approximation \cite{MCT}. This procedure 
gives the following
expression for the memory function:
\begin{equation}\label{Mmct}
M_{1 MCT}^{irr}(k;z) = \frac{nD_0}{V}\sum_{\mathbf{q}}
\left(\hat{\mathbf{k}}\cdot\mathbf{q} c(q)
\right)^2 S(q) F_1(|\mathbf{k}-\mathbf{q}|;z)
\end{equation}
Using Eq. (\ref{Mmct}) we can obtain a series expression for MCT's
time-integrated memory function.  We notice that 
$F_1(k;z=0) = (1+M_1^{irr}(k;z=0))/(D_0k^2)$ and iterate (\ref{Mmct})
to get
\begin{eqnarray}
&&M_{1 MCT}^{irr}(k=0;z=0)\label{Mmctseries}\nonumber
= \frac{n}{V}\sum_{\mathbf{q}_1} (\hat{\mathbf{k}}\cdot\mathbf{q}_1)^2
\frac{c(q_1)S(q_1)c(q_1)}{q_1^2}  \\ \nonumber
&& + \frac{1}{2}\left(\frac{n}{V}\right)^2 \sum_{\mathbf{q}_1,\mathbf{q}_2}
\left(\left(\hat{\mathbf{k}}\cdot\mathbf{q}_1\frac{1}{q_1^2}
\mathbf{q}_1\cdot\mathbf{q}_2
\right)^2 +\textrm{perm.}
\right)
\frac{c(q_1)S(q_1)c(q_1)c(q_2)S(q_2)c(q_2)}{|\mathbf{q}_1+\mathbf{q}_2|^2} 
\\ \nonumber 
&& + \frac{1}{3!}\left(\frac{n}{V}\right)^3
\sum_{\mathbf{q}_1,\mathbf{q}_2,\mathbf{q}_3}
\left(\left(\hat{\mathbf{k}}\cdot\mathbf{q}_1\frac{1}{q_1^2}
\mathbf{q}_1\cdot\mathbf{q}_2
\frac{1}{|\mathbf{q}_1+\mathbf{q}_2|^2}(\mathbf{q}_1+\mathbf{q}_2)
\cdot\mathbf{q}_3\right)^2
+\textrm{perm.}\right)\\ &&
\;\;\;\;\;\;\;\;\;\;\;\;\;\;\;\;\;\;\;\;\;\;\;\;\;\;\;\;\;\;\;\;\;\;\;         
\times
\frac{c(q_1)S(q_1)c(q_1)c(q_2)S(q_2)c(q_2)c(q_3)S(q_3)c(q_3)}
{|\mathbf{q}_1+\mathbf{q}_2+\mathbf{q}_3|^2} + \;\;\; ... ,
\end{eqnarray}
where we wrote the resulting series in a form similar to (\ref{Mseries}).
It should be noted that from each term of (\ref{Mseries}) 
the MCT expression (\ref{Mmctseries}) includes
only these terms that are explicitly positive definite.
We suggest that the difference between the two expressions
is the origin of MCT's well-known overestimation of so-called dynamic feedback 
effect. 

To investigate this last issue further we attempt to re-sum the
expression (\ref{Mseries}). To this end we resort to an 
approximation which simplifies angular dependence
of individual terms in series (\ref{Mseries}). This
approximation will be motivated in detail 
elsewhere \cite{long}. After the approximation we obtain
\begin{eqnarray}\label{Mseriesapp}
&& \sum_{\mathbf{q}_1,\mathbf{q}_2,...,\mathbf{q}_m}
\left(\hat{\mathbf{k}}\cdot\mathbf{q}_1\frac{1}{q_1^2}
\mathbf{q}_1\cdot\mathbf{q}_2
... \frac{1}{|\mathbf{q}_1+...+ \mathbf{q}_{m-1}|^2}
(\mathbf{q}_1+..+\mathbf{q}_{m-1})\cdot\mathbf{q}_m
+\textrm{perm.}\right)^2 \\ \nonumber &&
\times
\frac{c(q_1)S(q_1)c(q_1) ... c(q_m)S(q_m)c(q_m)}
{|\mathbf{q}_1+...+\mathbf{q}_m|^2} \approx
\frac{1}{d^m}\sum_{\mathbf{q}_1,\mathbf{q}_2,...,\mathbf{q}_m}
c(q_1)S(q_1)c(q_1) ... c(q_m)S(q_m)c(q_m),
\end{eqnarray}
where $d$ denotes dimensionality of the space.
One should note that Eq. (\ref{Mseriesapp}) reproduces exactly  
the first two terms in series (\ref{Mseries}). Also, Eq. 
(\ref{Mseriesapp}) is exact in one dimension.

Using (\ref{Mseriesapp}) we obtain 
the following simple formula for the mobile particle's diffusion coefficient: 
\begin{equation}\label{D}
D = \frac{D_0}{1+M_1^{irr}(k=0;z=0)} = 
D_0 \exp\left(- \frac{n}{d V}\sum_{\mathbf{q}} 
c(q)S(q)c(q)\right)
\end{equation}
It is clear that, for a generic interaction potential, the mobile particle's
motion is always diffusive; \textit{i.e.} the 
mobile particle is never localized.
One should also note a striking resemblance
between (\ref{D}) and Deem and Chandler's result \cite{DC} for the diffusion
coefficient of a single particle in Gaussian random media
(note that $-k_B T c(r)$ is the effective mobile 
particle-obstacle interaction potential \cite{effpot}). 
Comparison with numerical simulations of diffusion in Gaussian random media
showed that the latter result is extremely accurate.

In contrast, for a generic interaction potential, MCT's equation (\ref{Mmct}) 
(supplemented by the memory function representation for $F_1(k;z)$)
predicts that at sufficiently high obstacle density 
mobile particle's diffusion coefficient vanishes and the particle
gets localized.

It should be emphasized that the absence of the localization transition 
within the full theory and the resulting stark discrepancy 
with MCT is independent of
the approximation leading to Eq. (\ref{Mseriesapp}). 
The diffusion coefficient in $d$ dimensions is 
bounded from below by the $d=1$ diffusion coefficient \cite{JSP}. In 
one dimension the
series (\ref{Mseries}) can be re-summed without
difficulty. The result is given by formula (\ref{D}) with $d=1$.

\section{Final remarks} We showed here that, under 
the assumption of Gaussian factorization of 
static fluctuations, a single diffusing particle is never mobilized
by immobile obstacles. In contrast, MCT predicts localization 
at a high enough obstacle density.
One should notice that in some cases, 
\textit{e.g.} for a hard sphere
diffusing among immobile hard spheres, we expect the mobile
particle to get localized at sufficiently high obstacle density. 
Our result suggests that
this localization is intimately connected to non-Gaussian character 
of \emph{static} density fluctuations. Furthermore, MCT
implicitly assumes Gaussian factorization 
but it neglects
terms that cut off the localization transition. It would be of great 
interest to investigate whether in a more accurate theory these
terms are canceled by terms originating from non-Gaussian
character of static density fluctuations.
Finally, our results suggest that 
theories based on Gaussian density correlations
are not suitable for description of the glass transition.

\bigskip
ACKNOWLEDGMENT
\medskip
                           
I would like to thank David Reichman for a stimulating discussion.
Support by NSF Grant No. CHE 0111152 is gratefully acknowledged.


\begin{thebibliography}{99}
\bibitem{effpot} In the fully consistent approach
\cite{Gausfac} $-k_B T c(r) = V(r)$ and Deem and Chandler's result \cite{DC}
is recovered.
\end{thebibliography}
\end{document}